\documentclass[aps,prx,amsmath,nofootinbib,longbibliography,amssymb,superscriptaddress,twocolumn,10pt]{revtex4-2}

\usepackage[english]{babel}
\usepackage{graphicx}
\usepackage{ulem}
\usepackage{soul}
\usepackage{float}
\usepackage[svgnames]{xcolor}
\usepackage{comment}
\usepackage[colorlinks, linkcolor=NavyBlue , citecolor= NavyBlue,urlcolor = NavyBlue, breaklinks=true]{hyperref}
\usepackage{verbatim}
\usepackage{mathtools}
\usepackage{esint}
\usepackage{marginnote}

\newcommand{\defeq}{\vcentcolon=}

\newcommand{\RNum}[1]{\uppercase\expandafter{\romannumeral #1\relax}}

\def \tr {\textnormal{Tr}}
\begin{document}

\title{A Unifying Framework for Fractional Chern Insulator Stabilization}
\author{Peleg Emanuel}
\affiliation{Department of Condensed Matter Physics, Weizmann Institute of Science, Rehovot 76100, Israel}

\author{Anna Keselman}
\affiliation{Physics Department, Technion, 32000 Haifa, Israel}
\affiliation{The Helen Diller Quantum Center, Technion, 32000 Haifa, Israel}

\author{Yuval Oreg}
\affiliation{Department of Condensed Matter Physics, Weizmann Institute of Science, Rehovot 76100, Israel}

\date{\today}

\begin{abstract}
    We present a theory of fractional Chern insulator stabilization against charge-ordered states. We argue that the phase competition is captured by an effective interaction range, which depends on both the bare interaction range and quantum geometric properties. We claim that short effective interaction ranges stabilize fractional states while longer-range interactions favor charge-ordered states. To confirm our hypothesis, we conduct a numerical study of the generalized Hofstadter model using the density matrix renormalization group.
    Our theory offers a new interpretation of the geometric stability hypothesis and generalizes it, providing a unifying framework for several approaches to fractional phase stabilization.
    Finally, we propose a route towards experimental verification of the theory and possible implications for fractional states in bands with higher Chern numbers.
\end{abstract}

\maketitle

\textit{Introduction.---}
The discovery of the fractional quantum Hall effect (FQHE) \cite{tsui_two-dimensional_1982} marked a pivotal moment in modern condensed matter physics, opening new frontiers in the study of topological phases of matter and strongly interacting electron systems, and showing prospects for several technological applications \cite{nayak_non-abelian_2008}. Anyons, FQH quasiparticles with fractional charge and statistics, hold promise for enabling fault-tolerant quantum computation. Yet, the requirement of high magnetic fields for observing the FQHE has posed significant experimental challenges. This limitation spurred a theoretical exploration into lattice-based analogs of the FQHE, known as fractional Chern insulators (FCIs), for which no magnetic fields are needed \cite{kapit_exact_2010, qi_generic_2011, regnault_fractional_2011, sheng_fractional_2011, tang_high-temperature_2011, neupert_fractional_2011, liu_recent_2024}. Recently, FCIs have emerged as a rapidly developing area of research, with experimental breakthroughs demonstrating their existence in twisted MoTe$_2$ \cite{zeng_thermodynamic_2023, cai_signatures_2023, park_observation_2023} and rhombohedral pentalayer graphene \cite{lu_fractional_2024}, all under zero magnetic field.

Introducing a lattice allows for phases unachievable in the Landau levels (LLs) in which the FQHE has been observed. For example, while LLs have Chern number $\mathcal{C} = 1$, bands may generally have $|\mathcal{C}| > 1$, and could host novel fractional states \cite{liu_fractional_2012, sterdyniak_series_2013, moller_fractional_2015, andrews_stability_2018, andrews_stability_2021}. Nevertheless, searching for material candidates in the plethora of lattice systems is challenging.

In a seminal work \cite{roy_band_2014}, Roy suggested conditions under which an isolated flat band mimics the lowest Landau-level (LLL) and hence could support FQHE ansatzes \textit{mutatis mutandis}. The conditions relate to the quantum geometry of the band, commonly characterized by the Fubini-Study metric $g_{ij} \left(\boldsymbol{k}\right)$, and Berry curvature $\Omega \left(\boldsymbol{k}\right)$ (for definitions, see Appendix \ref{appendix: quantum geometry}).
If these quantities are uniform in the Brillouin zone, and further satisfy the trace condition $\tr g = |\Omega|$,
the projected density operators satisfy the GMP algebra \cite{girvin_collective-excitation_1985, girvin_magneto-roton_1986}. As a result, the problem may be mapped onto the LLL \cite{parameswaran_fractional_2012, roy_band_2014}.

Since, it has been shown that the uniformity of quantum geometry is not crucial for FCI stabilization \cite{varjas_topological_2022,  morales-duran_pressure-enhanced_2023, wang_origin_2023}.
On the other hand, the trace condition $\tr g = |\Omega|$  has been shown to support LL generalizations \cite{claassen_position-momentum_2015, wang_exact_2021, wang_origin_2023}, and to be a special case of \textit{vortexability} \cite{ledwith_vortexability_2023}, a band-property suggesting Laughlin states \cite{laughlin_anomalous_1983} are energetically favorable.
Because realistic systems rarely satisfy equality conditions, minimizing $\tr g - |\Omega|$ has been suggested to promote FCI stability, a hypothesis often referred to as the \textit{geometric stability hypothesis} (GSH) \cite{jackson_geometric_2015}.
The hypothesis has been shown to hold numerically \cite{parker_field-tuned_2021, morales-duran_pressure-enhanced_2023, jackson_geometric_2015}, far from the saturation of the trace inequality \cite{shavit_quantum_2024}, and in the absence of an LL limit \cite{bauer_fractional_2022, andrews_stability_2024}.
Experimental verification of the GSH is a standing challenge due to the difficulty of probing quantum geometric properties in the lab.

Despite its continued success, the GSH presents an incomplete picture. Even though FCIs are strongly interacting states, interaction plays little role in the theory since the quantum geometric properties on which it relies are band properties.
To justify analyzing a single isolated flat band, one often requires a hierarchy of energy scales, $W \ll V \ll E_g$, with $W$ and $E_g$ the bandwidth and the band gap, and $V$ an energy scale associated with the interaction \cite{bergholtz_topological_2013}. However, the \textit{form} of the interaction, namely, its dependence on spatial coordinates, plays no role in the GSH. 
For comparison, the spatial extent of the interaction is key for analyzing FQH stability, most notably through the lens of the celebrated Haldane pseudopotentials \cite{haldane_fractional_1983}.

\begin{figure}[!t]
    \centering
    \includegraphics[width=1.0\linewidth]{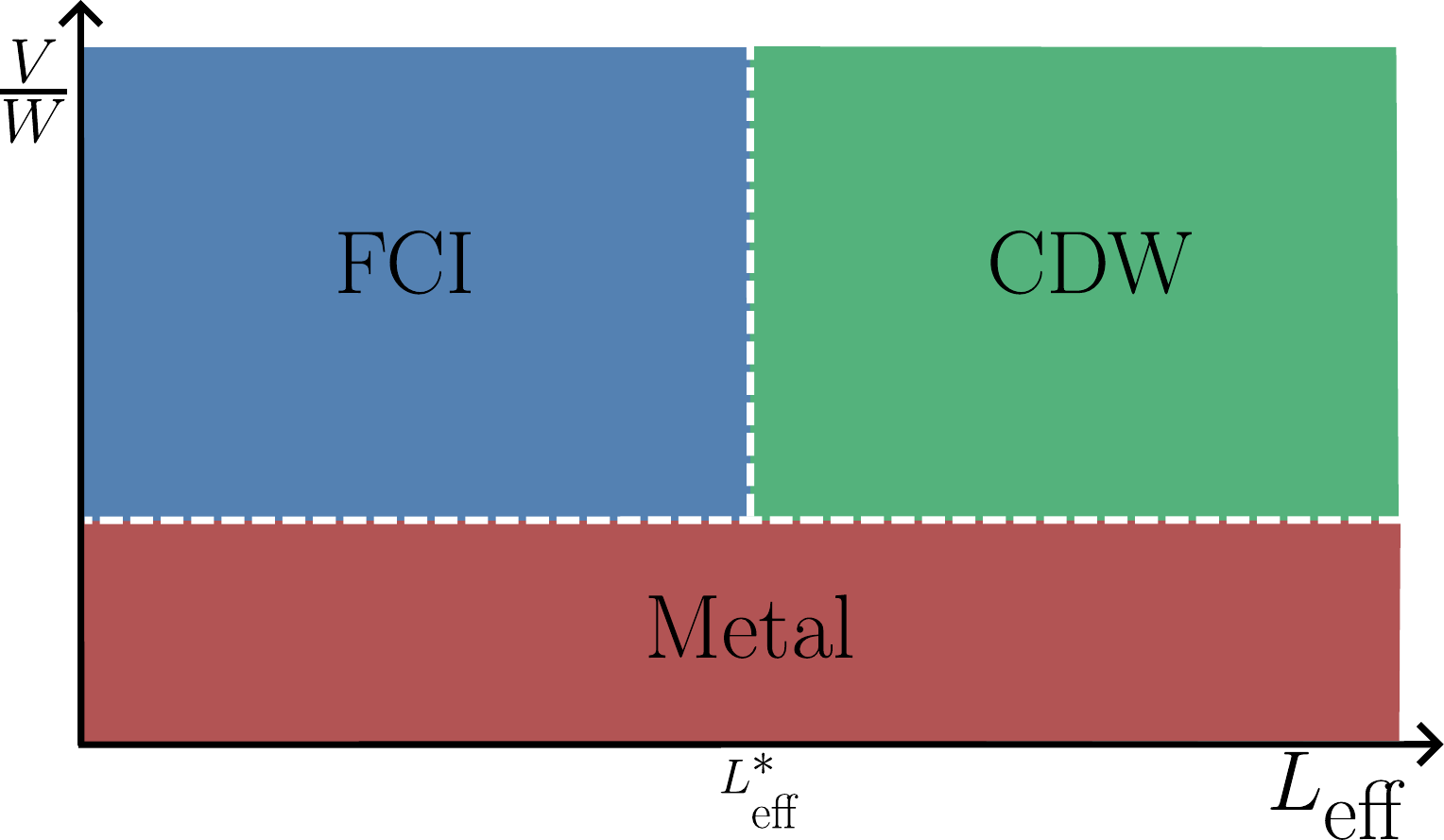}
    \caption{Proposed qualitative phase diagram. $V$ is the interaction energy scale, $W$ the bandwidth, and $L_{\textnormal{eff}}$ the effective interaction range. FCI and CDW stand for fractional Chern insulator and charge density wave, respectively. Phase boundaries present trends with $L_{\textnormal{eff}}$ and $V/W$ and do not necessarily run parallel to the axes. The transition length, $L_{\textnormal{eff}}^*$, is not universal.}
    \label{fig:qualitative}
\end{figure}

In this letter, we propose a generalization of the GSH. 
Since $\tr g$ measures wavefunction spread, projected interactions are extended by larger $\tr g$, leading to a larger effective interaction range, $L_{\textnormal{eff}}$.
We argue that transitions between fractional states and charge-ordered states are controlled by $L_{\textnormal{eff}}$,
as seen in the proposed qualitative phase diagram of Fig.~\ref{fig:qualitative} (an additional metallic phase is depicted for completeness).
$L_{\textnormal{eff}}$ increases with $\tr g$, in agreement with the GSH,
but also depends on the bare interaction range, assigning a role to the form of interactions.
In all that follows, we assume the bare interaction is short-ranged, such that a bare interaction range $L_\textnormal{bare}$ is well defined.

Contrary to the conventional formulation of the GSH, our theory stresses the importance of minimizing $\tr g$ rather than $\tr g - |\Omega|$.
Minimizing either quantity is equivalent if the Berry curvature doesn't change sign.
However, as we elaborate in the discussion section, the two conditions differ in their predictions for bands with $|\mathcal{C}| > 1$ {and more numerical evidence is required to tell which of the two is a better FCI indicator}.
Finally, we suggest possible ways to test our theory and utilize it to stabilize fractional states in the lab.

\textit{Qualitative Picture.---}
To start, we analyze the success of the Laughlin wavefunction as a ground-state Ansatz.
The spatial extent of the interaction is essential, as the following intuition suggests.
The LLL wavefunctions are analytic up to a Gaussian normalization factor. Due to fermion statistics, any two-body wavefunction in the LLL must be proportional to $\left(z_1 - z_2\right)^\alpha$, with $\alpha$ odd and $z_j = x_j + i y_j$ the complex coordinate of electron $j$.
Consequently, the short-range pair probability scales as $r^{2\alpha}$ with $r$ the particle distance. For the $1/3$ filling Laughlin state, $\alpha=3$, so for short enough distances the pair probability falls below that of any $\alpha=1$ state. Therefore, for sufficiently short-ranged interactions, the Laughlin state is lower in energy.
This idea is more formally captured by short-ranged artificial interactions known as pseudopotentials \cite{haldane_fractional_1983, trugman_exact_1985, ledwith_fractional_2020, ledwith_strong_2021, ledwith_vortexability_2023}.

In the opposite limit of interaction ranges much longer than the magnetic length $\ell_B$, short-range correlations matter little in the calculation of the energy. In comparison, long-range correlations, displayed by charge-ordered states, could reduce it considerably.
We thus expect that fractional state stability is reduced if the interaction length scale becomes large enough, compatible with numerical results \cite{kupczynski_interaction-driven_2021, lu_continuous_2025}.

The form of the wavefunction may greatly alter the effective interaction length scale. Consider, for example, interacting electrons occupying the $N$\textsuperscript{th}~LL. Screening from occupied lower LLs leads to a decrease in the bare interaction range, $L_{\textnormal{bare}}$, with the level index. Nevertheless, Fogler at al. \cite{fogler_ground_1996} show the level-projected interaction is roughly given by $\sim \Theta \left(2R_C - |x|\right)$ for large $N$, with $R_C$ the cyclotron radius, equal to $\sqrt{2N + 1}\ell_B$, and $\Theta(x)$ the Heaviside step function. The effective interaction range, $L_{\textnormal{eff}}$, thus increases with $N$, explaining the observation of fractional states for $N=0,1$ only, and charge-ordered states known as bubbles and stripes for $N \ge 2$ \cite{fogler_laughlin_1997, fogler_stripe_2002}.

In high LLs, the wavefunction spatial extent sets the range of interaction. Simply put, states that overlap spatially interact more strongly. Generally, $\tr{g}$ is a good indicator for the wavefunction spread \cite{verma_quantum_2025}, and indeed, for LLs, $\sqrt{\tr g} = R_c$ \cite{ozawa_relations_2021}. 
Thus, the above picture intuitively explains the GSH. The smaller $g$ is, the more localized the wavefunctions, and hence the effective interactions, are. As a result, the Laughlin state admits a lower energy. To stabilize Laughlin-like states, one can either modify the quantum geometry of a band to reduce $\tr g$, or the bare interaction to reduce $L_\textnormal{bare}$.

\begin{figure}[!t]
    \centering
    \includegraphics[width=0.9\linewidth]{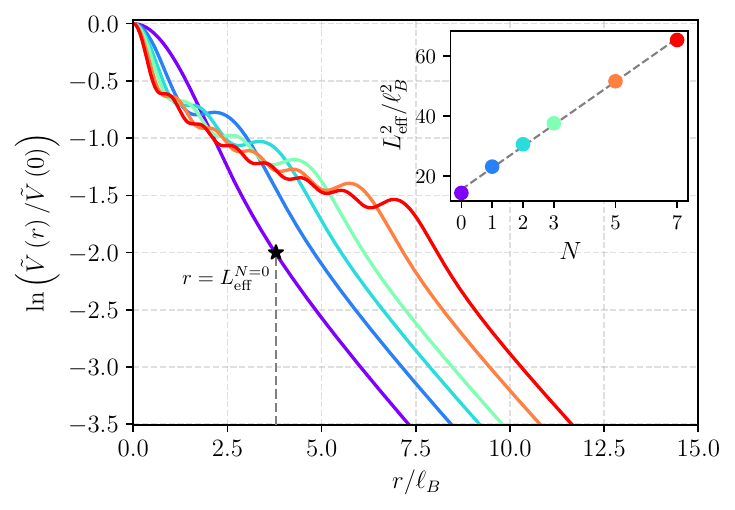}
    \caption{Effective interactions in real space for different LLs. The bare interaction is taken to be $V_q \propto \tanh{qd} / q$ with $d = 5\ell_B$, motivated by screening in double-gated devices.
    In the inset, $L_{\textnormal{eff}}$ vs. LL index $N$, defined such that $\ln \left(\tilde V\left(L_{\textnormal{eff}}\right) / \tilde V\left(0\right)\right) = -2$. The dashed gray line presents a linear fit of $L_{\textnormal{eff}}^2 \left(N\right)$, as suggested by the scaling of $\sqrt{\tr g} \sim \sqrt{N}$.}
    \label{fig:landau_scale}
\end{figure}

The emergence of a Landau-level dependent effective length scale is illustrated in Fig. \ref{fig:landau_scale}.
The effective interactions in the $N$\textsuperscript{th} Landau level are given by
$\tilde V _q = 
V_q \left(L_N\left(q^2 \ell_B ^2 / 2\right)\right)^2 \exp \left(-q^2 \ell_B ^2 / 2\right)$, $V_q$ being bare interaction and $L_{N}$ the $N$\textsuperscript{th} Laguerre polynomial \cite{fogler_stripe_2002}. Oscillations may be observed at shorter distances, followed by non-oscillatory decay at longer distances, with an effective decay length obeying $L_\textnormal{eff} \propto \sqrt{\tr g}$. A similar behavior of $L_\textnormal{eff}$ has been observed in Ref.~\cite{may-mann_how_2025}. 

The conclusions obtained for LLs may be naturally generalized to lattice systems.
Consider an isolated spin-polarized electron band with density-density interactions. We further assume the band is relatively flat, and neglect dispersion, such that the Hamiltonian is given by
\begin{equation}
    \mathcal{H} = \frac{1}{2A}\sum_{\boldsymbol{q} \neq 0} V_{q} \sum_{\boldsymbol{k} \boldsymbol{k'}}
    \Lambda _{\boldsymbol{k}} \left(\boldsymbol{q}\right) \Lambda _{\boldsymbol{k'}} \left(-\boldsymbol{q}\right)
    c^\dagger _{\boldsymbol{k} + \boldsymbol{q}}
    c^\dagger _{\boldsymbol{k'} - \boldsymbol{q}}
    c_{\boldsymbol{k'}}
    c_{\boldsymbol{k}}
    ,
    \label{eq: flat-band hamiltonian}
\end{equation}
with $c^\dagger$ ($c$) fermionic creation (annihilation) operators, $V_{q}$ an isotropic interaction strength in momentum space, $A$ the system area and 
$\Lambda _{\boldsymbol{k}} \left(\boldsymbol{q}\right) = \langle u_{\boldsymbol{k} + \boldsymbol{q}} | u_{\boldsymbol{k}} \rangle$ the form factors.{ 
If $|\Lambda _{\boldsymbol{k}} \left(\boldsymbol{q}\right)|$ are $\boldsymbol{k}$-independent, as they are in LLs, we may similarly define an effective interaction $\tilde V_q = V_q |\Lambda _{\boldsymbol{k}} \left(\boldsymbol{q}\right)|
|\Lambda _{\boldsymbol{k'}} \left(\boldsymbol{-q}\right)|$, with an associated length scale $L_\textnormal{eff}$. $L_\textnormal{eff}$ may be easily identified by examining $\tilde V$ in real space, as is illustrated in Fig.~\ref{fig:landau_scale} for Landau levels.

Since $\Lambda_{\boldsymbol{k}} \left(0\right) = 1$ and commonly $V_q$ peaks at $q = 0$, 
we assume small~$q$ transfer processes dominate and neglect higher $\boldsymbol{q}$ terms. We may now expand the form factors, 
$|\Lambda _{\boldsymbol{k}} \left(\boldsymbol{q}\right)| = 1 - \frac{1}{2} g_{ij} \left(\boldsymbol{k}\right) q^i q^j + \mathcal{O} \left(q^4\right)$
with $g_{ij} \left(\boldsymbol{k}\right)$ the Fubini-Study metric \cite{cheng_quantum_2013, ledwith_strong_2021}, and $i,j =x,y$. In leading order in $\boldsymbol{q}$, denoting $\theta_{\boldsymbol{k}} \left(\boldsymbol{q}\right) = \arg \Lambda _{\boldsymbol{k}} \left(\boldsymbol{q}\right)$,

\begin{gather}
    \label{eq: appx hamiltonian}
    \begin{aligned}
        \mathcal{H} \approx \frac{1}{2A}\sum_{\boldsymbol{q} \neq 0} V_{q} \sum_{\boldsymbol{k} \boldsymbol{k'}}
        &\left(
            \frac{e ^{i \left(\theta_{\boldsymbol{k}} \left(\boldsymbol{q}\right) + \theta_{\boldsymbol{k'}} \left(-\boldsymbol{q}\right)\right)}}{1 + \frac{1}{2}\left(
            g_{ij} \left(\boldsymbol{k}\right) + g_{ij} \left(\boldsymbol{k'}\right)
        \right) q^i q^j}
        \right)
        \\ & \times c^\dagger _{\boldsymbol{k} + \boldsymbol{q}}
        c^\dagger _{\boldsymbol{k'} - \boldsymbol{q}}
        c_{\boldsymbol{k'}}
        c_{\boldsymbol{k}}
        .
    \end{aligned}
\end{gather}

This approximation allows us to study the effect of the Fubini-Study metric on the effective interaction, which may be written as $\tilde V_q \approx V_q / \left(1 + g_{ij} q^iq^j\right)$.
By the convolution theorem $\tilde V\left(\boldsymbol{r}\right) \approx V\left(\boldsymbol{r}\right) * \mathcal{F}^{-1} \left[\left(1 + g_{ij} q^i q^j\right)^{-1}\right]$, with $\mathcal{F}^{-1}$ the inverse Fourier transform. Hence, the effective interaction is smoother and more extended in real space than the bare one.
The above statements hold as long as $g$ is $\boldsymbol{k}$-independent, otherwise, the obtained $L_\textnormal{eff}$ is $\boldsymbol{k}$- and $\boldsymbol{k'}$-dependent.
Yet, the same picture should \textit{qualitatively} hold in the more general case. The Fubini-Study metric reshapes the effective interaction, with larger $g$ leading to smoother, more spread interactions.

To sum it up,  we hypothesize FCI stability is determined by an effective interaction length scale $L_{\textnormal{eff}}$ that increases with both $L_{\textnormal{bare}}$ and $g$, with larger $L_{\textnormal{eff}}$ promoting a competing charge order.
In LLs, charge order manifests as stripes or bubbles that break continuous translation symmetry. Studying lattice systems that do not possess this continuous symmetry, we consider the more general charge density waves (CDWs). CDWs are often considered as competing phases to FCIs \cite{parker_field-tuned_2021, shavit_quantum_2024, morales-duran_pressure-enhanced_2023, grushin_enhancing_2012, wilhelm_interplay_2021, liu_recent_2024, aronson_displacement_2025,
abouelkomsan_band_2024,he_fractional_2025}, especially at high filling fractions~\cite{wilhelm_interplay_2021}.

\textit{Model.---}
To test our hypothesis, we attempt to recreate FCI-CDW phase transitions by modifying either $\tr g$ or $L_\textnormal{bare}$. If the transition is determined by $L_{\textnormal{eff}}$ that increases with both $L_{\textnormal{bare}}$ and $g$, as we claim, the bare transition length $L^*_{\textnormal{bare}}$ should decrease with $\tr g$.

We study the generalized Hofstadter model \cite{andrews_hofstadtertools_2024, hofstadter_energy_1976} on a square lattice. Unless hopping parameters are fine-tuned, the model bands approach LLs as flux is decreased \cite{bauer_fractional_2022, andrews_stability_2024}, suggesting FCI-CDW transitions could be observed in it. Furthermore, by modifying the ratio between hopping parameters, we can tune $\tr g$ \cite{andrews_stability_2024, andrews_hofstadtertools_2024}. The Hamiltonian is given by

\begin{equation}
    \mathcal{H} = -\!\!\!\sum_{l} t_l
    \sum_{\langle ij\rangle_l} \left(e^{i\theta_{ij}} c^\dagger _i c_j
    +h.c.\right) + V\!\!\!\sum_{l \le n_{\textnormal{NN}}} \sum_{\langle ij\rangle_l} n_i n_j
    ,
    \label{eq: gen-hof}
\end{equation}
where $n_i = c^\dagger _i c_i$ is the number operator, and $\langle ij \rangle_l$ denote~$l$ nearest neighbors. The phases $\theta_{ij}$ are set by a Peierls substitution for a uniform flux, as detailed in Appendix~\ref{appendix: genhof}. In all that follows, we take $t_1 = 1$, $t_l = 0$ for~$l > 2$.

$n_{\textnormal{NN}}$ sets the interaction range. For simplicity, we consider flat interactions, with a coefficient $V = V_0 / A_{\textnormal{int}}$ where $A_{\textnormal{int}} = \sum_{l \le n_{\textnormal{NN}}} \# \langle ij \rangle_l$ is the number of sites with which each site interacts, and $V_0$ a constant energy scale. The continuum definition of this interaction is given by $V\left(r\right) = \left(V_0 / \pi R^2\right) \Theta \left(R - r\right)$, with $R$ the interaction range. In momentum space, $V\left(q\right) = 2V_0 J_1\left(Rq\right) / Rq$, with $J_1$ the Bessel function of the first kind. It is peaked at $\boldsymbol{q} = 0$, justifying the approximation of Eq.~\ref{eq: appx hamiltonian}.

With 3 unit cells per unit flux, 3 Hofstadter minibands develop. At $t_2 = 0$ their Chern numbers are $C = 1, -2, 1$ \cite{dana_quantised_1985, andrews_hofstadtertools_2024} and will not change as long as $t_2$ is small enough such that the gap doesn't close. In all that follows, we populate the system such that the lowest miniband could be at $1/3$ filling, namely, $L_x L_y / 9$ electrons, with $\left(L_x, L_y\right)$ the system dimensions. Note that we do not perform any subspace projection. It has been recently observed that the FQH ground state has support on higher LLs and that increasing LL mixing promotes charge order \cite{teng_solving_2024, abouelkomsan_band_2024,haug_interaction-driven_2025}. Hence, it is crucial to account for multiple bands in our system.

\textit{Results.---}
We perform an infinite density matrix renormalization group (iDMRG) study of the system in the cylinder geometry using {\ttfamily TeNPy} \cite{hauschild_efficient_2018, hauschild_tensor_2024}.
To identify topological order, we thread flux through the cylinder and calculate the Hall conductivity \cite{zaletel_flux_2014, grushin_characterization_2015}. Exemplary flux threading results displaying $\sigma_H = 0, 1/3$ are presented in Fig.~\hyperref[fig: main results]{3b}. 
As a measure of charge order, we calculate the magnitude of the maximal Bragg peak in the structure factor $S_{\boldsymbol{q}}$. For more details, see Appendix \ref{appendix: phase indicators}.

\begin{figure*}[t!]\includegraphics[width=1.0\linewidth]{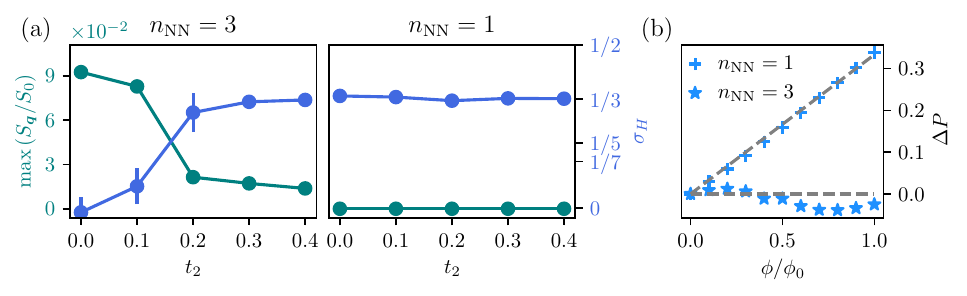}
    \caption{(a) Hall conductivities and maximal Bragg peaks in the ground state for different values of $t_2$ and $n_{\textnormal{NN}}$ obtained using iDMRG. Calculated with $V_0 \left(t_2 = 0\right) = 30$ and a unit cell of dimensions $\left(24, 9\right)$. (b) Charge pumping for $t_2 = 0$ and, and $n_{\textnormal{NN}} = 1, 3$. $\Delta P$ is the charge polarization relative to the $\phi = 0$ ground state, $\phi$ is the threaded flux and $\phi_0$ the flux quantum. In gray, slopes of $0$ and $1/3$ for comparison.
    }
    \label{fig: main results}
\end{figure*}

To assess the effects of $L_{\textnormal{eff}}$, we modify $n_{\textnormal{NN}}$ and $t_2$, setting $L_{\textnormal{bare}}$ and $\tr g$, respectively. We note that $t_2$ determines multiple band properties that possibly affect FCI stability. In the range $t_2 \in \left[0, 0.5\right]$, as $t_2$ increases, $\tr g$ decreases, the lowest miniband narrows, and the gap to the next miniband increases, as detailed in Table \ref{tab:genhof_params}. Hence, a larger $t_2$ is expected to promote fractional phases, though not necessarily due to quantum geometric effects. We now discuss means of isolating the effects of $\tr g$.

Small band gaps might endanger FCI stability by inducing non-negligible support of the ground state on other bands. This effect may be checked for a posteriori by calculation of the projection of the ground state onto the lowest miniband, as detailed in Appendix \ref{appendix: band projection}. 
If the ground state lies mostly within a single band, two energy scales remain in the system: the interaction scale $V_0$ and the bandwidth $W$. The dependence of the FCI-CDW transition on $W$ should thus be through the dimensionless ratio $V_0/ W$. Therefore, to isolate the effect of the Fubini-Study metric, we perform calculations with a $V_0$ that is $t_2$ dependent and scales with the bandwidth, such that the ratio $V_0\left(t_2\right) / W\left(t_2\right)$ is constant. 

Results of iDMRG calculations for $n_{\rm NN} = 1, 3$ and different values of $t_2$ are presented in Fig.~\hyperref[fig: main results]{3a}~\cite{resultsNote}. The bond dimension used is $\chi=1500$ for calculations of the ground state and $\chi=1000$ for flux threading, resulting in a maximal truncation error of less than $10^{-6}$. 
For $n_{\textnormal{NN}} = 3$, loss of topological response is accompanied by a sharp increase in Bragg peak magnitude as $t_2$ is lowered below~$0.2$, which we identify as an FCI-CDW transition.
Projections onto the lowest miniband are above $95\%$ for all values of $n_{\textnormal{NN}}$ and $t_2$.
CDW ground state projections are slightly lower, $95.1\%$ and $96.8\%$ for $t_2 = 0, 0.1$, respectively, at $n_{\textnormal{NN}} = 3$. In comparison, all fractional states have a lowest-miniband support of over $98\%$.

Remarkably, the isolated band description holds despite the isolated band condition $V \ll E_g$ being far from satisfied.
Generally, form factors lead to differences between inter- and intra-band interaction terms, and may promote the occupation of higher bands if $V \gtrsim E_g$. In our system, $V_0$ exceeds $E_g$, yet higher miniband projections are extremely low.
A possible explanation lies in the suppression of off-diagonal form factors due to small $\boldsymbol{q}$ dominance. This could reduce energy gains from occupying higher bands, as detailed in Appendix \ref{appendix: effectively isolated bands}.

The high projections observed rule out major effects of the single-particle gap and indicate changes in $\tr g$ are the ones that drive the FCI-CDW transition for $n_{\textnormal{NN}} = 3$. The absence of the transition for $n_{\textnormal{NN}} = 1$ shows the $\tr g$ does not determine it alone, $L_\textnormal{bare}$ must exceed a critical length scale $L_\textnormal{bare}^*$. Finally, 
we find that $L_\textnormal{bare} ^*$ decreases with $\tr g$, affirming a key prediction of our theory.

\textit{Discussion.---}
We present a new theory of FCI stabilization, suggesting an effective interaction scale $L_{\textnormal{eff}}$ controls the transition between fractional states and charge density waves, unifying existing approaches.
We interpret the geometric stability hypothesis, explaining how quantum geometric properties renormalize interactions, modifying $L_{\textnormal{eff}}$ such that the Laughlin state is a more likely ground state. However, $L_{\textnormal{eff}}$ doesn't depend on geometry alone. In addition, our theory agrees with other well-established approaches for the stability of fractional states \cite{haldane_fractional_1983, fogler_laughlin_1997, fogler_stripe_2002, kupczynski_interaction-driven_2021}.

We showed that it is possible to cross the FCI-CDW transitions by modifying either $\tr g$, adding to existing numerical evidence in support of the GSH, or $L_{\textnormal{bare}}$, compatible with recent observations \cite{kupczynski_interaction-driven_2021, lu_continuous_2025}.
In addition, $L_{\textnormal{bare}}^*$ decreases with $\tr g$, in agreement with our interpretation of the transition being controlled by $L_{\textnormal{eff}} \left(L_{\textnormal{bare}}, g\right)$, unifying the two approaches.

Although we have been able to show that increasing $L_{\textnormal{eff}}$, leads to a transition out of the FCI phase, we have yet to establish the effective length at which the transition occurs, $L_{\textnormal{eff}} ^*$.
We stress that due to the $\boldsymbol{k}$-dependence of quantum geometric quantities, it is generally impossible to uniquely define $L_{\textnormal{eff}}$. Similarly, rather than providing an explicit expression for $L_{\textnormal{eff}}^*$, we aim to understand its qualitative dependence on different band properties.
In lattice systems, there are generally several quantities of dimension length, including lattice constants, $\sqrt{\tr g_{\boldsymbol{k}}}$, and $\sqrt{|\Omega_{\boldsymbol{k}}|}$, all of which possibly affect $L_{\textnormal{eff}} ^*$.
$\boldsymbol{k}$-dependence of quantum geometric properties could also introduce additional relevant scales.

Understanding the qualitative behaviour of $L_\textnormal{eff}^*$ could help identify relevant FCI indicators. For instance,
whether or not $L_{\textnormal{eff}} ^*$ depends on $\Omega$
has implications for the stabilization of fractional states in higher Chern bands. Since $|\Omega|$ bounds $\tr g$ from below, $|\mathcal{C}| > 1$ bands will generally have higher $\tr g$ and thus larger $L_{\textnormal{eff}}$. If $L_{\textnormal{eff}} ^*$ does not increase accordingly, our theory suggests it is harder to stabilize FCIs in higher Chern bands. 
Differently put, if $L_{\textnormal{eff}} ^*$ does not increase with $|\Omega|$, our theory stresses the importance of minimizing $\tr g$, rather than the commonly defined trace indicator $\tr g - |\Omega|$. Usually, minimizing either quantity is roughly equivalent, so most existing numerical evidence, ours included, does not serve to differentiate between the two. To deduce whether $\tr g$ or $\tr g - |\Omega|$ is a better indicator for FCI stability, one would need to isolate the effects of the Chern number.

The special role attributed to the Fubini-Study metric is due to the leading order expansion of the form factors. This approximation is often justified for Coulomb interaction and its screened variants, highly peaked at $\boldsymbol{q} = 0$, confirming its physical relevance. Nonetheless, this result is not universal, and it might be required to expand the form factor beyond second-order or consider it in its entirety.
The definition of $L_{\textnormal{eff}}$ does not require the above approximation. 
Recall $\tilde V$ is defined using the entire form factors (see discussion below Eq.~\ref{eq: flat-band hamiltonian}). Thus, our results allow the extension of the GSH to regimes in which form factor expansion is unjustified, and terms that are independent of the quantum metric need to be taken into account.

We note that wavefunction localization can also reduce the energy of Wigner crystals, and has recently been suggested to promote charge order \cite{haug_interaction-driven_2025}. Our argument is based on the dominance of the short-range part of the interaction, and we expect it to hold in the dense limit, where the wavefunction spread exceeds the inter-particle distance and significant overlap is inevitable. In the opposite limit, interaction may be dominated by its inter-particle distance component, and our claim breaks down, possibly explaining the discrepancy.

Our hypothesis opens up a route for experimental verification and could possibly aid in stabilizing FCIs.
Generally, it is hard to modify quantum geometric properties in the lab without inducing numerous other changes in the system. However, there are other means to change the form of effective interactions, such as screening.
Typically, screening has little impact on short-distance behavior while reducing interaction at longer distances. Hence, the resulting interaction is more concentrated at short distances and better supports FCIs. An important caveat is that screening reduces interaction strength altogether, possibly promoting a metallic phase if the interaction energy scale is too small.
A similar suggestion has recently appeared for the stabilization of fractional topological insulators \cite{kwan_when_2024}.

The numerical evidence supporting our theory was obtained by studying the generalized Hofstadter model, which, though famed for its simplicity and versatility, does not serve to describe materials in which FCIs have been experimentally observed.
Recent progress in the application of neural networks in the studies of fractional states \cite{teng_solving_2024, luo_solving_2025} opens the path to numerical studies of more realistic models, and could perhaps help assess our proposal for stabilizing FCIs in the lab.

\textit{Acknowledgements.---}
We thank Yaar Vituri for fruitful discussions.
This research was funded in part by the DFG Collaborative Research Center (CRC) 183, and by ISF grants No. 1914/24 and No. 2478/24.
A.K. acknowledges funding by the ISF grant No. 2443/22.

\appendix

\section{Quantum Geometry}
\label{appendix: quantum geometry}
We briefly overview the fundamentals of quantum geometry for single bands in two dimensions. Quantities defined below may be extended to multi-band systems as well as manifolds other than the Brillouin zone. We refer the reader to Refs. \cite{verma_quantum_2025, yu_quantum_2025, provost_riemannian_1980, cheng_quantum_2013, ozawa_relations_2021} for a more general discussion.

Let $\psi\left(\boldsymbol{k}\right)$ be the periodic Bloch wavefunction of a gapped band, with $\boldsymbol{k}$ the lattice momentum. To measure how $\psi$ changes as one moves in the Brillouin zone, one would naively suggest the distance $ds^2 = \langle \partial_i \psi | \partial_j \psi\rangle dk^i dk^j$. However, it is not $U\left(1\right)$-gauge invariant, prompting the definition of the quantum geometric tensor (QGT) \cite{provost_riemannian_1980, cheng_quantum_2013},
\begin{equation}
    Q_{ij} \left(\boldsymbol{k}\right)= \langle \partial_i \psi | \left(1 - | \psi \rangle \langle \psi |\right) |\partial_j \psi \rangle.
\end{equation}
$Q\left(\boldsymbol{k}\right)$ is hermitian and may be divided into a real symmetric part $g_{ij} \left(\boldsymbol{k}\right) = \Re Q_{ij} \left(\boldsymbol{k}\right)$ and an anti-symmetric imaginary part, $\Im Q_{ij} \left(\boldsymbol{k}\right) = \frac{1}{2} \epsilon_{ij} \Omega \left(\boldsymbol{k}\right)$ with $\Omega$ the Berry curvature and $\epsilon$ the antisymmetric tensor. The Fubini-Study metric $g$ defines the gauge invariant infinitesimal distance $ds^2 = g_{ij} dk^i dk^j$.
For every $\boldsymbol{k}$, $\tr g \left(\boldsymbol{k}\right) \ge |\Omega \left(\boldsymbol{k}\right)|$, an inequality commonly known as the \textit{trace inequality} \cite{roy_band_2014}. Its saturation, $\tr g \left(\boldsymbol{k}\right) = |\Omega \left(\boldsymbol{k}\right)|$, is known as the \textit{trace condition}.

The QGT may also be 
defined using the position operators $\hat r_i$,
$Q_{ij} = \langle \psi |\hat r_i \left(1- |\psi \rangle \langle \psi |\right) \hat r_j | \psi \rangle$. Evidently, it is similar in form to the second moment of the position operator. In fact, it is its gauge invariant part and therefore sets a lower bound on Wannier localization \cite{marzari_maximally_1997, marzari_maximally_2012}. The above intuitively explains why $\tr g$ is a good measure for the wavefunction spread \cite{verma_quantum_2025, yu_quantum_2025}.

\section{Generalized Hofstadter model}
\label{appendix: genhof}

The dynamics of a particle in a uniform magnetic field on a square grid, with $n$-nearest neighbor hopping terms, are governed by the Hamiltonian in Eq.~\ref{eq: gen-hof}, with the phases,
\begin{equation}
    \theta_{ij} = \frac{2\pi}{\phi_0}\int _{\boldsymbol{r_i}} ^{\boldsymbol{r_j}} d\boldsymbol{r} \cdot \boldsymbol{\mathcal{A}}\left(\boldsymbol{r}\right)
    ,
\end{equation}
where $\phi_0$ is the flux quantum and $\boldsymbol{\mathcal{A}}$ the magnetic vector potential.
We consider a family of linear gauges, $\boldsymbol{\mathcal{A}} = -B_x y \boldsymbol{\hat x} + B_y x \boldsymbol{\hat y}$, such that $B_x + B_y = B$, which include the commonly used symmetric and Landau gauges. Straight line trajectories of $n_\alpha$ steps in direction $\alpha$ may be parametrized as $\boldsymbol{r} \left(\eta\right) = \boldsymbol{r}_0 + \eta a\left(n_x \boldsymbol{\hat x} + n_y \boldsymbol{\hat y}\right)$, with $\eta \in \left[0, 1\right]$. For $\left(n_x, n_y\right) = \left(x_j - x_i, y_j - y_i\right) / a$, with $a$ the lattice constant,
\begin{equation}
    \theta_{ij} = 2\pi \frac{p}{q}
        \left(
            \left(n_y \frac{B_y}{B} \frac{x_i}{a} - n_x \frac{B_x}{B} \frac{y_i}{a}\right)
            + \frac{1}{2}n_x n_y \frac{B_y - B_x}{B}
        \right)
\end{equation}
with $p$ unit fluxes per $q$ unit cells. 

Selected parameters for values of $t_2$ from Fig.~\hyperref[fig: main results]{3a} appear in Table~\ref{tab:genhof_params}. Dispersions and distributions of quantum geometric properties for representative values of $t_2$ appear in Fig.~\ref{fig:band_properties}. Energies are given in units of $t_1$ and quantum geometric properties in units of $a^2$.

\begin{figure}[!b]
    \centering
    \includegraphics[width=1.0\linewidth]{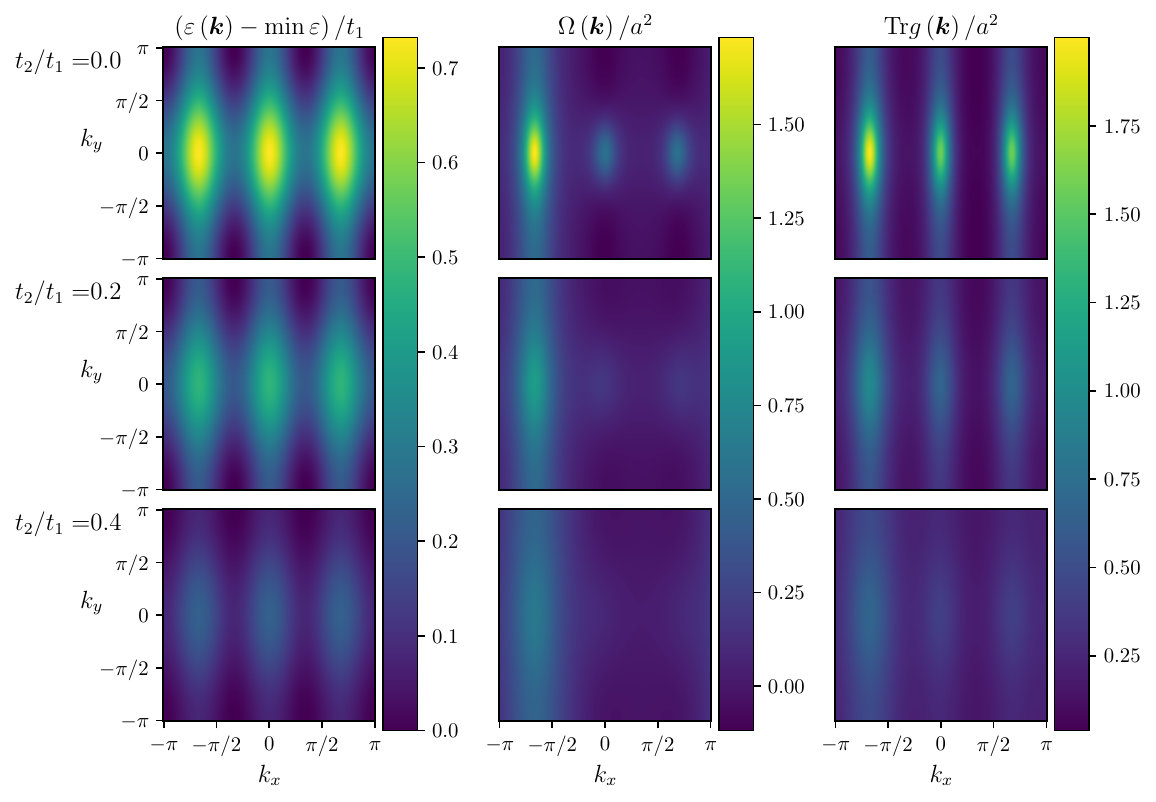}
    \caption{Band properties of the generalized Hofstadter model as a function of the lattice momentum $\boldsymbol{k}$ for $t_2/t_1 = 0.0, 0.2, 0.4$, calculated in the Landau x gauge ($B_y = B$).}
    \label{fig:band_properties}
\end{figure}

\begin{table}[!t]
    \centering
    \begin{tabular}{||c c c c||} 
        \hline
        $t_2 / t_1$ & $ \langle \tr g \rangle /a^2$ & $W / t_1$ & $E_g / t_1$ \\ [0.5ex] 
        \hline\hline
        0.0 & 0.366 & 0.732 & 1.27 \\ 
        \hline
        0.1 & 0.337 & 0.605 & 1.74 \\
        \hline
        0.2 & 0.322 & 0.478 & 2.21 \\
        \hline
        0.3 & 0.315 & 0.352 & 2.68 \\
        \hline
        0.4 & 0.313 & 0.225 & 3.16 \\ [1ex]
        \hline
    \end{tabular}
    \caption{Parameters of the lowest band of the generalized Hofstadter model for different values of $t_2/t_1$.}
    \label{tab:genhof_params}
\end{table}

\section{Calculations of phase indicators}
\label{appendix: phase indicators}

\begin{figure*}[t!]
    \centering
    \includegraphics[width=1.0\linewidth]{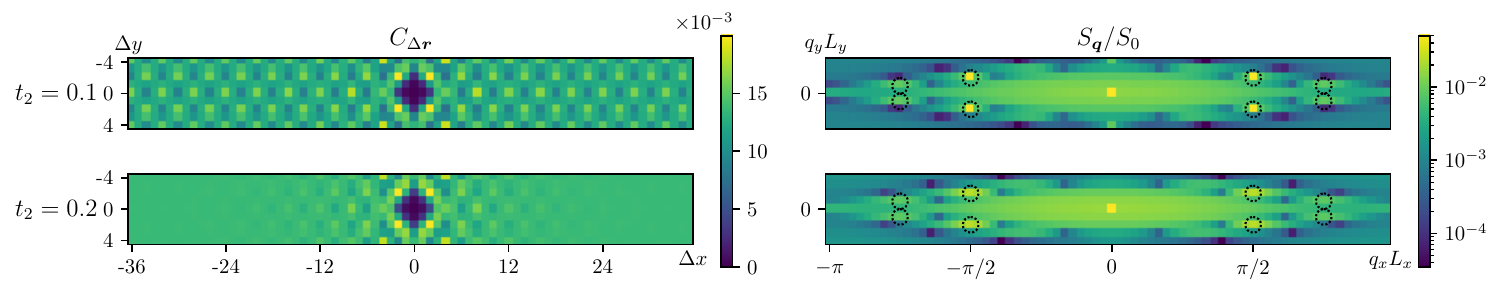}
    \caption{Charge order indicators for $t_2 = 0.1, 0.2$ and $n_\textnormal{NN} = 3$ of Fig. \ref{fig: main results}. Left, Average density-density correlation, calculated over 3 DMRG unit cells. Right, corresponding normalized structure factors. Found Bragg peaks are circled.}
    \label{fig:structure factor}
\end{figure*}

\subsection{Hall response}
To calculate Hall response, we increasingly thread flux through the system, with a step of size $\Delta \phi = 0.1\phi_0$. For each value of $\phi$, we calculate the charge polarization from the Schmidt decomposition \cite{zaletel_flux_2014}, allowing us to calculate charge transfer per step $j$, $\Delta c^j$, and the Hall response $\sigma_H ^{j} = \Delta c^j / \Delta \phi_0$. The values displayed in Fig.~\hyperref[fig: main results]{3a} are the average $\Delta c^j$, with error bars given by the standard deviation.
Note that the system might jump between degenerate ground states as we thread flux through the system. Hence, we omit the two largest and two smallest $\Delta c$ from the above analysis.

\subsection{Bragg peaks}

We calculate the average density-density correlator
\begin{equation}
    C_{xy} \defeq \frac{1}{A}\sum_{x_0, y_0} \langle n_{x + x_0, y+ y_0} n_{x_0, y_0} \rangle,
\end{equation}
over 3 DMRG unit cells, and the structure factor by its 2-dimensional Fourier transform, $S_{\boldsymbol{q}} = \mathcal{F} \left[C\right]$.
We calculate the maximal Bragg peak in $S_{\boldsymbol{q}} / S_0$, excluding $\boldsymbol{q} = 0$ and possible peaks at $q_\alpha = \pm \pi / L_{\alpha}$ for $\alpha = x, y$. Exemplary correlations and structure factor as well as obtained Bragg peaks are displayed in Fig.~\ref{fig:structure factor}, both in the CDW phase ($t_2 = 0.1$) and the FCI phase ($t_2 = 0.2$).

\subsection{Band projection}
\label{appendix: band projection}
The projector onto the $\alpha$ band is given by $\mathcal{P}_\alpha = \sum_{\boldsymbol{k}} | \psi_{\alpha\boldsymbol{k}} \rangle \langle \psi_{\alpha\boldsymbol{k}} |$, with $\boldsymbol{k}$ the lattice momenta, and $| \psi_{\alpha\boldsymbol{k}} \rangle$ the Bloch wavefunctions. Bloch's theorem implies $| \psi_{\alpha\boldsymbol{k}} \rangle = \frac{1}{\sqrt{A}} \exp \left(i\boldsymbol{k} \cdot \boldsymbol{r}\right)| u_{\alpha \boldsymbol{k}} \rangle$, with $| u_{\alpha \boldsymbol{k}} \rangle$ the periodic Bloch wavefunctions and $A$ the system size. It is convenient to expand the states in real space, in which the iDMRG result is given, and write $|u_{\alpha \boldsymbol{k}} \rangle = \sum_i u_{\alpha \boldsymbol{k} i} |i\rangle$, where $i$ indicates both real space and orbital degrees of freedom. The projection of a wavefunction $|\phi \rangle$ is thus

\begin{gather}
\begin{aligned}
    \langle \phi |\mathcal{P}_\alpha |\phi \rangle
    & = \frac{1}{A}\sum_{ij} \sum_{\boldsymbol{k}} e^{i \boldsymbol{k} \cdot \left(\boldsymbol{r_i} - \boldsymbol{r_j}\right)}
    u_{\boldsymbol{k} i} u_{\boldsymbol{k} j} ^*\langle \phi | i \rangle \langle j | \phi \rangle
    \\ & = \frac{1}{A} \sum_{ij} \sum_{\boldsymbol{k}} e^{i \boldsymbol{k} \cdot \left(\boldsymbol{r_i} - \boldsymbol{r_j}\right)}
    u_{\boldsymbol{k} i} u_{\boldsymbol{k} j} ^* \langle c^\dagger _i c_j \rangle
\end{aligned}
\end{gather}
with $\boldsymbol{r_i}$ the spatial coordinate at $i$. The expectation values $\langle c^\dagger _i c_j \rangle$ are straightforward to obtain from a matrix product state, and the coefficients may be calculated analytically, diagonalizing the non-interacting part of the Hamiltonian in Eq.~\ref{eq: gen-hof}. Note that for iDMRG, one, in principle, needs to calculate expectation values over infinitely many unit cells. We perform calculations with ten.

\section{Effectively isolated bands}
\label{appendix: effectively isolated bands}
As mentioned in the main text, lowest miniband projections are very high despite the isolated flat band condition not being satisfied.
Hence, there's no apriori justification for Eq.~\ref{eq: flat-band hamiltonian}, and we should consider the more general form \cite{ledwith_strong_2021}
\begin{widetext}
    \begin{equation}
        \mathcal{H} = \sum_{\boldsymbol{k}} \sum_\alpha \varepsilon_{\boldsymbol{k} \alpha} c_{\boldsymbol{k} \alpha} ^\dagger c_{\boldsymbol{k} \alpha}
        + \frac{1}{2A}\sum_{\boldsymbol{q}} V_{q} \sum_{\alpha \beta \gamma \delta} \sum_{\boldsymbol{k} \boldsymbol{k'}}
        \Lambda^{\alpha \beta} _{\boldsymbol{k}} \left(\boldsymbol{q}\right) \Lambda^{\gamma \delta} _{\boldsymbol{k'}} \left(-\boldsymbol{q}\right)
        : c^\dagger _{\alpha \boldsymbol{k} + \boldsymbol{q}}
        c_{\beta \boldsymbol{k}}
        c^\dagger _{\gamma \boldsymbol{k'} - \boldsymbol{q}}
        c_{\delta \boldsymbol{k'}}
        :,
    \label{eq: general hamiltonian}
    \end{equation}
\end{widetext}
where $::$ denotes normal ordering, Greek letters denote band indices and
$\Lambda ^{\alpha\beta}_{\boldsymbol{k}} \left(\boldsymbol{q}\right) \defeq 
\langle u^\alpha _{\boldsymbol{k} + \boldsymbol{q}} | u^\beta _{\boldsymbol{k}} \rangle$ are the interband form factors.

The periodic Bloch wavefunctions form an orthonormal basis for every lattice momentum and therefore $\sum_{\beta} |\Lambda ^{\alpha\beta}_{\boldsymbol{k}} \left(\boldsymbol{q}\right)|^2 = 1$. 
Assuming small momentum transfer processes dominate, we expand in small $q$ to leading order and find, for $\beta \neq \alpha$, $|\Lambda ^{\alpha\beta}_{\boldsymbol{k}} \left(\boldsymbol{q}\right)| \le \sqrt{g^\alpha _{ij} \left(\boldsymbol{q}\right) q^i q^j} \le \sqrt{\tr g ^\alpha} |q|$. We approximate the problem by omitting $\beta \neq \alpha$ and $\delta \neq \gamma$ terms and take the diagonal form factors to be $\sim 1$ for relevant values of $\boldsymbol{q}$. Thus, form factors don't serve to differentiate significantly inter- and intra-band interactions. Possible energy gains from occupying the next miniband rather than the lowest are expected to be much smaller than $V_0$, possibly not exceeding the gap even if $V_0$ does. This could explain the observed high lowest miniband projections.


\begin{thebibliography}{69}
\bibitem [{res()}]{resultsNote}%
  \BibitemOpen
  \href@noop {} {}\bibinfo {note} {Results for $n_{\textnormal{NN}} = 2$ are almost identical to those of $n_{\textnormal{NN}} = 1$, and were therefore omitted from the paper. Results for $n_{\textnormal{NN}} = 4$ presented considerable finite size effects, which is expected for $n_{\textnormal{NN}} \sim L_y / 2$. These persisted for $L_y=11$. Hence, we only consider $n_{\textnormal{NN}} \le 3$.}\BibitemShut {Stop}%
\end{thebibliography}
\end{document}